\documentstyle[amssymb,preprint,aps]{revtex}

\begin{document}
\title{${\cal PT-}$ symmetric pseudo-perturbation recipe; an imaginary cubic
oscillator with spikes.}
\author{\ {\bf Omar Mustafa}}
\address{Department of Physics, Eastern Mediterranean University, \\
G. Magosa, North Cyprus, Mersin 10, Turkey\\
email:omar.mustafa@emu.edu.tr}
\author{{\bf Miloslav Znojil}}
\address{Department of Theoretical Physics, Institute of Nuclear Physics, \\
Academy of Sciences, 25068 \v{R}e\v{z}, Czech Republic (mailing address)\\
and\\
Doppler Institute of Mathematical Physics, Fac. Nucl. Sci. and Phys. Eng.,\\
Czech Technical University, 115 19 Prague, Czech \ Republic\\
email:znojil@ujf.cas.cz\\
PACS: 03.65.Fd, 03.65.Ge.}
\date{\today}
\maketitle

\begin{abstract}
{\small \ The pseudo-perturbative shifted - }$\ell $ {\small expansion
technique PSLET is shown applicable in the non-Hermitian }${\cal PT\ }$%
{\small - symmetric context. The construction of bound states for several }$%
{\cal PT\ }${\small - symmetric potentials is presented, with special
attention paid to }$V(r)=ir^{3}-\alpha \sqrt{ir}$ {\small oscillators.}
\end{abstract}

\newpage

\section{Introduction}

In their recent studies Dorey, Dunning and Tateo (DDT) [1] have considered
the manifestly non-Hermitian Schr\"{o}dinger equation, in $\hbar =2m=1$
units, 
\begin{equation}
\left[ -\,\frac{d^{2}}{dr^{2}}+\frac{\ell (\ell +1)}{r^{2}}-\alpha \,\sqrt{%
i\,r}+i\,r^{3}\right] \,{\large \psi }_{k,\ell }(r)=E_{k,\ell }{\large \psi }%
_{k,\ell }(x).
\end{equation}
They have rigorously proved that the spectrum $E_{k,\ell }$ is real and
discrete in the domain of the sufficiently large angular momenta, 
\begin{equation}
\ell >\max \left[ \frac{1}{4}(2\alpha -7),-\frac{1}{2}\right] \equiv \ell
_{DDT}(\alpha ).
\end{equation}
This inspired our subsequent study of this model [2] where we have shown
that in the strong coupling regime with $\ell \gg 1$, the low lying DDT
bound states may be very well approximated by the harmonic oscillators. At
the same time we have noticed that the quality of such an asymptotic
approximation may deteriorate quite significantly with both the increase of
excitation $k$ and/or with the decrease of $\ell$.

Such a situation is, obviously, challenging. Firstly, our study [2] revealed
that the manifest non-Hermiticity of the models of the type (1) leads to the
reliable leading order approximation {\em only after} we select our harmonic
oscillator approximant as lying {\em very far} from the real axis (i.e.,
from the Hermitian regime). Such a recipe is, apparently, deeply
incompatible with a smooth modification of the traditional zero-order
approximants occurring in current Hermitian $1/\ell$ recipes (cf. a small
sample of some references in [3]). At the same time, the smallness of $%
1/\ell $ still supports the feeling that the similar perturbation techniques 
{\em should} prove efficient after their appropriate modification.

This observation offered a sufficiently strong motivation for our continued
interest in the complex, non-Hermitian model (1) which may be understood as
a characteristic representative of a very broad class of the so called
pseudo-Hermitian models with real spectra, the analyses of which became very
popular in the recent literature [4] - [18]. Within this class, the
strong-coupling version of DDT oscillators (1) with $\alpha \gg 1$ forms a
particularly suitable testing ground as it combines the necessary reality of
its spectrum with the smallness of the inverse quantity $1/\ell $. Moreover,
the phenomenologically appealing non-Hermitian models like (1) are rarely
solvable in closed form so that the presence of a ''universal" small
parameter $1/\ell \ll 1$ offers one of not too many ways towards their
systematic approximate solution.

In our present paper II we intend to discuss such a possibility in more
detail.

\section{Framework}

The first stages of interest in the non-Hermitian oscillators (1) date back
to an old paper by Caliceti et al [5]. It studied the imaginary cubic
problem in the context of perturbation theory and, more than twenty years
ago, it offered the first rigorous explanation why the spectrum in such a
model may be real and discrete. In the literature, this result has been
quoted as a mathematical curiosity [6] and only many years later, its
possible relevance in physics re-emerged and has been emphasized [7]. This
initiated an extensive discussion which resulted in the proposal of the so
called ${\cal PT}$ symmetric quantum mechanics by Bender et al [8].

The key idea of the new formalism lies in the empirical observation that the
(phenomenologically desirable) existence of the real spectrum need not
necessarily be attributed to the Hermiticity of the Hamiltonian. The current
Hermiticity assumption $H=H^{\dagger }$ is replaced by the mere ${\cal PT}$
symmetry $H=H^{\ddagger }\equiv {\cal PT}H{\cal PT}$. Here, ${\cal P}$
denotes the parity (${\cal P}x{\cal P}=-x$) while the anti-linear operator $%
{\cal T}$ mimics the time reflection (${\cal T}i{\cal T}=-i$). It is easy to
verify that example (1) exhibits such a type of symmetry [2] and may serve
as an elementary illustration of the latter extension of quantum mechanics.

The Bender's and Boettcher's conjecture that $H=H^{\ddagger }$ may imply $%
{\rm Im}\,E=0$ is fragile. The extent as well as limitations of its validity
are most easily analyzed in the language of linear algebra using the
biorthogonal bases [4,9] and/or exactly solvable Hamiltonians [10].
Nevertheless, the relevance of many unsolvable oscillators originates from
their applicability in physics [11] and field theory [12]. In such a setting
it is necessary to develop and test also some efficient approximation
methods. New and intensive studies employed the ideas of the strong-coupling
expansions [13] as well as the complex version of WKB [14], Hill
determinants and Fourier transformation [15], functional analysis [16], \
variational and truncation techniques [17], and linear programming [18].

In what follows we intend to use the method based on the smallness of the
inverse angular momentum parameter $1/\ell $. Various versions [19] of such
an approach are available for Hermitian models where the combination of the
central repulsive core $\ell (\ell +1)/r^{2}$ with a confining (i.e.,
asymptotically growing) interaction $V(r)$ forms a practical effective
potential $V_{p}(r)$ which possesses a pronounced minimum. Near such a
minimum the shape of the potential is naturally fitted by the elementary and
solvable harmonic oscillator well. Corrections can be then evaluated in an
unambiguous and systematic manner~[20].

As long as we intend to move to the complex plane, the leading-order
approximation may become non-unique. One finds several different complex
and/or real minima of $V_{p}(r)$ even in our oversimplified examples (1)
[2]. For all the similar non-Hermitian Hamiltonians, even the most
sophisticated forms of the perturbation expansions in the powers of our
small parameter $1/\ell$ lose their intuitive background and deserve careful
new tests, therefore.

Some of the related reopened questions will be clarified by our forthcoming
considerations inspired, basically, by the pseudo-perturbative shifted$-\ell$
expansion technique (PSLET) in its form designed for the standard Hermitian
Hamiltonians and described, say, by Mustafa and Odeh in ref. [20]. In this
recipe, the shifts of $\ell$ were admitted as suitable optional auxiliary
parameters while the use of the prefix ''pseudo-" just indicated that $\ell$
itself is an artificial, kinematical parameter rather than a genuine
dynamical coupling.

\section{${\cal PT}$-symmetric PSLET recipe}

As we already mentioned, one of the first ${\cal PT-}$symmetric models with
an immediate impact on physics has been the Buslaev's and Grecchi's quartic
anharmonic oscillator [7] described by the radial Schr\"{o}dinger equation
in $d-$dimensional space,

\begin{equation}
\left[ -\frac{d^{2}}{dr^{2}}+\frac{\ell _{d}~(\ell _{d}+1)}{r^{2}}+V(r)%
\right] \chi _{k,\ell }(r)=E_{k,\ell }~\chi _{k,\ell }(r).
\end{equation}
In this model they shifted the coordinate axis to the complex plane, $r=t-ic$
with a constant $\ \Im (r)=-c<0$ and variable $\Re (r)=t$ $\in (-\infty
,\infty )$. They also required that $\chi _{k,\ell }(r)\in L_{2}(-\infty
,\infty )$ at all the partial waves $\ell _{d}=\ell +(d-3)/2$ and dimensions 
$d>2$.

This example may find the various sophisticated generalizations some of
which will be also mentioned in due course in what follows. For example, a $%
t-$dependent shift $c=c(t)$ may be needed both for the exactly solvable
Coulombic model of ref.[21] and for all the more general and purely
numerically tractable potentials $V(r)\sim -(ir)^{N}$of ref.[8] with the
positive exponents $N>3$. Fortunately, the transition to $c=c(t)$ remains
particularly elementary, being mediated by the mere change of the variable
in eq.(3) within this class (c.f., e.g., ref.[21] for an explicit
illustration). Another remark might concern the assumption that the wave
functions are square integrable. For the most elementary ${\cal PT-}$
symmetric Hamiltonians this assumption seems very natural but in certain
more sophisticated models its use may require a more careful analysis as
presented, e.g. in refs. [4] and [16].

The practical experience with the Hermitian version of the
pseudo-perturbation shifted - $\ell $ expansion technique of Mustafa and
Odeh [20] may serve as a key inspiration of an appropriate complexified new
PSLET recipe. Firstly we notice a formal equivalence between the assumed
smallness of our parameter $1\,/\,\ell _{d}\thickapprox 0$ and of its (
arbitrarily) shifted form. Thus we introduce a new symbol $\bar{l}=\ell
_{d}-\beta $ and, simultaneously, move and rescale our coordinates $%
r\longrightarrow x$ to \newline
\begin{equation}
x=\bar{l}^{1/2}(r-r_{0}).
\end{equation}
Here $r_{0}$ is an arbitrary point, with its particular value to be
determined later. Equation (3) thus becomes

\begin{equation}
\left\{ -\bar{l}~\frac{d^{2}}{dx^{2}}+\frac{\bar{l}^{2}+(2\beta +1)~\bar{l}%
+\beta ~(\beta +1)}{r_{0}^{2}~\left[ x/(r_{0}~\bar{l}^{1/2})+1\right] ^{2}}+%
\frac{\bar{l}^{2}}{Q}~V(x(r))\right\} \Psi _{k,\ell }(x(r))=E_{k,\ell }\Psi
_{k,\ell }(x(r)),
\end{equation}
where $Q$ is a constant that scales the potential $V$ at large-$\ell _{d}$
limit and is set, for any specific choice of $\ell _{d}$ and $k,$ equal to $%
\bar{l}^{2}$ at the end of calculation. Expansions about this point, $x=0$
(i.e. $r=r_{0}$), yield

\begin{equation}
\frac{1}{r_{0}^{2}~\left[ x/(r_{0}\bar{l}^{1/2})+1\right] ^{2}}%
=\sum_{n=0}^{\infty }~(-1)^{n}~\frac{(n+1)}{r_{0}^{n+2}}~x^{n}~\bar{l}%
^{-n/2},
\end{equation}

\begin{equation}
\frac{\bar{l}^{2}}{Q}~V(x(r))=\sum_{n=0}^{\infty }~\left( \frac{d^{n}V(r_{0})%
}{dr_{0}^{n}}\right) ~\frac{x^{n}}{n!\hspace{-0.02in}~Q}~\bar{l}^{-(n-4)/2}.
\end{equation}
It is also convenient to expand $E_{k,\ell }$ as

\begin{equation}
E_{k,\ell }=\sum_{n=-2}^{\infty }E_{k,\ell }^{(n)}~\bar{l}^{-n}.
\end{equation}
Of course one may consider also energy coefficient of half-entire power of \ 
$\bar{l}$ in (8) but all these coefficients vanish ( c.f., e.g.,
ref.[19,20]). Equation (5), therefore, reads 
\begin{eqnarray}
&&{\Huge \{}-\frac{d^{2}}{dx^{2}}+\sum_{n=0}^{\infty }B_{n}~x^{n}~\bar{l}%
^{-(n-2)/2}+(2\beta +1)\sum_{n=0}^{\infty }T_{n}~x^{n}~\bar{l}^{-n/2} 
\nonumber \\
&&+\beta ~(\beta +1)\sum_{n=0}^{\infty }T_{n}~x^{n}~\bar{l}^{-(n+2)/2}~%
{\Huge \}}\Psi _{k,\ell }(x)  \nonumber \\
&=&\left\{ \sum_{n=-2}^{\infty }E_{k,\ell }^{(n)}~~\bar{l}^{-(n+1)}\right\}
~\Psi _{k,\ell }(x)
\end{eqnarray}
\newline
where

\begin{equation}
~B_{n}=T_{n}+\left( \frac{d^{n}V(r_{0})}{dr_{o}^{n}}\right) ~\frac{1}{n!~Q}%
~;~~T_{n}=(-1)^{n}~\frac{(n+1)}{r_{0}^{n+2}}.
\end{equation}
Equation (9) is to be compared with the non-Hermitian ${\cal PT}$\ \ -
symmetrized perturbed harmonic oscillator in the one dimensional
Schr\"{o}dinger equation \newline
\begin{equation}
\left[ -\frac{d^{2}}{dy^{2}}+\frac{\omega ^{2}}{4}~\left( y-ic\right)
^{2}+\varepsilon _{0}+P(y-ic)\right] \Phi _{k}(y)=\lambda _{k}~\Phi _{k}(y),
\end{equation}
where $P(y-ic)$ is a complexified perturbation - like term and $\varepsilon
_{0}$ is obviously a constant. Such a comparison implies 
\[
\varepsilon _{0}=B_{0}\,\,\bar{l}+(2\beta +1)T_{0}+\beta (\beta +1)T_{0}\,/\,%
\bar{l} 
\]
\begin{eqnarray}
&&\lambda _{k}=\varepsilon _{0}+(2k+1)\frac{\omega }{2}+\sum_{n=0}^{\infty
}\lambda _{k}^{(n)}~\bar{l}^{-(n+1)}  \nonumber \\
\,\,\,\quad &=&B_{0}\,\,\bar{l}+{\Large [}(2\beta +1)T_{0}+(2k+1)\frac{%
\omega }{2}{\Large ]+}\frac{1}{\,\bar{l}}\left[ \beta (\beta
+1)T_{0}+\lambda _{k}^{(0)}\right] +\sum_{n=2}^{\infty }\lambda _{k}^{(n-1)}~%
\bar{l}^{-n}  \nonumber \\
\quad \quad \quad \ \ \ \ \ &=&E_{k,\ell }^{(-2)}~~\bar{l}+E_{k,\ell
}^{(-1)}+\sum_{n=1}^{\infty }E_{k,\ell }^{(n-1)}~\bar{l}^{-n}
\end{eqnarray}

The first two dominant terms are obvious\newline
\begin{equation}
E_{k,\ell }^{(-2)}=\frac{1}{r_{0}^{2}}+\frac{V(r_{0})}{Q},
\end{equation}
\newline
\begin{equation}
E_{k,\ell }^{(-1)}=\frac{(2\beta +1)}{r_{0}^{2}}+(2~k+1)~\frac{\omega }{2},
\end{equation}
and with appropriate rearrangements we obtain\newline
\begin{equation}
E_{k,l}^{(0)}=\frac{\beta (\beta +1)}{r_{0}^{2}}+\lambda _{k}^{(0)},
\end{equation}

\begin{equation}
E_{k,\ell }^{(n)}=\lambda _{k}^{(n)}~~;~~~~n\geq 1.
\end{equation}
Here $r_{0}$ is chosen to minimize $E_{k,\ell }^{(-2)}$, i. e.

\begin{equation}
\frac{dE_{k,\ell }^{(-2)}}{dr_{0}}=0~~~~and~~~~\frac{d^{2}E_{k,\ell }^{(-2)}%
}{dr_{0}^{2}}>0.
\end{equation}
Equation (13) in turn gives, with $~(\ell _{d}-\beta )^{2}=Q$,

\begin{equation}
|~\ell _{d}-\beta ~|=\sqrt{\frac{r_{0}^{3}V^{^{\prime }}(r_{0})}{2}}~.
\end{equation}
Consequently, $B_{0}\,\bar{l}\,($ $=\bar{l}~E_{k,\ell }^{(-2)})$ adds a
constant to the energy eigenvalues and $B_{1}=0.$ The next leading
correction to the energy series, $\bar{l}~E_{k,\ell }^{(-1)}$, consists of a
constant term and the exact eigenvalues of the unperturbed one-dimensional
harmonic oscillator potential $\omega ^{2}x^{2}/4\,(=B_{2}\,x^{2}),$ where 
\begin{equation}
0<\omega =\omega ^{\left( \pm \right) }=\pm ~\frac{2}{\ r_{0}^{2}}~\Omega
~;~~\Omega =\sqrt{3+\frac{r_{0\ }V^{^{\prime \prime }}(r_{0})}{V^{^{\prime
}}(r_{0})}}~,
\end{equation}
Evidently, equation (19) implies that $r_{0}$ is either pure real, $\omega
=\omega ^{\left( +\right) },$ or pure imaginary, $\omega =\omega ^{\left(
-\right) }$. Next, the shifting parameter $\beta $ is determined by choosing 
$\bar{l}~E_{k,\ell }^{(-1)}=0$. That is 
\begin{equation}
\beta =\beta ^{\left( \pm \right) }=-\frac{1}{2}\left[ 1\pm (\ 2\ k+1)\
\Omega {}\right] ,
\end{equation}
\newline
where $\beta =\beta ^{\left( +\right) }$ \ for $r_{0}$ pure real and $\beta
=\beta ^{\left( -\right) }$ \ for $\ r_{0}$ pure imaginary. Then equation
(9) reduces to

\begin{equation}
\left[ -\frac{d^{2}}{dx^{2}}+\sum_{n=0}^{\infty }v^{(n)}~\bar{l}^{-n/2}%
\right] ~\Psi _{k,\ell }(x)=\left[ \sum_{n=-1}^{\infty }E_{k,\ell }^{(n)}~%
\bar{l}^{-(n+1)}\right] ~\Psi _{k,\ell }(x).
\end{equation}
\newline
with 
\begin{equation}
v^{(0)}(x)=B_{2}~x^{2}+(2~\beta +1)~T_{0},
\end{equation}
\newline
and for $n\geq 1$ 
\begin{equation}
v^{(n)}(x)=B_{n+2}~x^{n+2}+(2~\beta +1)~T_{n}~x^{n}+\beta ~(\beta
+1)~T_{n-2}~x^{n-2}.
\end{equation}
Equation (21) upon setting the wave functions \newline
\[
\Psi _{k,\ell }(x)=F_{k,\ell }(x)~exp(U_{k,\ell }(x)) 
\]
\newline
readily transforms into the Riccati-type equation:\newline
\begin{eqnarray*}
&&F_{k,\ell }(x)\left[ -\left( U_{k,\ell }^{^{\prime \prime }}(x)+U_{k,\ell
}^{^{\prime }}(x)U_{k,\ell }^{^{\prime }}(x)\right) +\sum_{n=0}^{\infty
}v^{(n)}(x)\bar{l}^{-n/2}\right. \\
&&\left. -\sum_{n=0}^{\infty }E_{k,\ell }^{(n-1)}\bar{l}^{-n}\right]
-2F_{k,\ell }^{^{\prime }}(x)U_{k,\ell }^{^{\prime }}(x)-F_{k,\ell
}^{^{\prime \prime }}(x)=0,
\end{eqnarray*}
\newline
where the primes denote derivatives with respect to $x$. It is evident that
this equation admits solutions (c.f., e.g., ref.[20]) of the form \newline
\[
U_{k,\ell }^{^{\prime }}(x)=\sum_{n=0}^{\infty }U_{k}^{(n)}(x)~~\bar{l}%
^{-n/2}+\sum_{n=0}^{\infty }G_{k}^{(n)}(x)~~\bar{l}^{-(n+1)/2}, 
\]
\newline
\[
F_{k,\ell }(x)=x^{k}+\sum_{n=0}^{\infty
}\sum_{p=0}^{k-1}a_{p,k}^{(n)}~~x^{p}~~\bar{l}^{-n/2}, 
\]
\newline
with\newline
\[
U_{k}^{(n)}(x)=\sum_{m=0}^{n+1}D_{m,n,k}~~x^{2m-1}~~~~;~~~D_{0,n,k}=0, 
\]
\newline
\[
G_{k}^{(n)}(x)=\sum_{m=0}^{n+1}C_{m,n,k}~~x^{2m}. 
\]
\newline
Obviously, equating the coefficients of the same powers of \ $\bar{l}$ and $%
x $ ( for each $k$), respectively, one can calculate the energy eigenvalues
and eigenfunctions ( following the uniqueness of power series
representation) from the knowledge of $C_{m,n,k}$, $D_{m,n,k}$, and $%
a_{p,k}^{(n)}$ in a hierarchical manner.

In order to test the performance of our strategy, let us first apply it to
the two trivial, exactly solvable ${\cal PT}$- symmetric examples.

\section{An elementary illustration of the recipe}

\subsection{${\cal PT-}$symmetric Coulomb}

Using the potential $V(r)=iA/r$ (where $A$ is a real coupling constant) in
the above ${\cal PT-}$symmetric PSLET setting, one reveals the leading-order
energy approximation 
\begin{equation}
\bar{l}^{2}E_{k,\ell }^{(-2)}=\frac{\bar{l}^{2}}{r_{0}^{2}}+\frac{iA}{r_{0}}.
\end{equation}
The unique minimum at $r_{0}=2i\bar{l}^{2}/A$ occurs in the upper-half of
the complex plane. In this case $\Omega =1,~\beta =\beta ^{(-)}=k,$ the
leading energy term reads 
\begin{equation}
\bar{l}^{2}E_{k,\ell }^{(-2)}=\frac{iA}{2r_{0}}=\frac{A^{2}}{4(\ k-\ell
_{d}\ )^{2}},
\end{equation}
and higher-order corrections vanish identically. Therefore, the total energy
is 
\begin{equation}
E_{n,\ell }=\frac{A^{2}}{4(\ n-2\ \ell -1)^{2}},~~n=1,2,3,...
\end{equation}
where $n=k+\ell +1$ is the principle quantum number. Evidently, the
degeneracy associated with ordinary ( Hermitian) Coulomb energies $%
E_{n}=-A^{2}/(2n)^{2}$ is now lifted upon the complexification of, say, the
dielectric constant embedded in $A$. Moreover, the phenomenon of {\em flown
away states} at $k=\ell _{d}$ emerge, of course if they do exist at all ( i.
e. the probability of finding such states is presumably zero, the proof of
which is already beyond our current methodical proposal). Therefore, for
each $k-$ state there is an $\ell _{d}-$ state to {\em fly away}.

Next, let us replace the central-like repulsive/attractive core through the
transformation $\ell _{d}(\ell _{d}+1)\rightarrow \alpha _{o}^{2}-1/4$, i.e. 
$\ell _{d}=-1/2+q|\alpha _{o}|,$ with $q=\pm 1$ denoting {\em quasi-parity},
and recast (25) as 
\begin{equation}
E_{k,q}=\frac{A^{2}}{(2k+1-2q|\alpha _{o}|)^{2}}.
\end{equation}
Which is indeed the exact result obtained by Znojil and L\'{e}vai [21].
Equation (27) implies that {\em even-quasi-parity}, $q=+1,$ states with $%
k=|\alpha _{o}|-1/2$ \ {\em fly away \ }and disappear from the spectrum.
Nevertheless, {\em quasi-parity-oscillations }\ are now manifested by energy
levels crossings. That is, a state-$k$ with {\em even-quasi-parity} crosses
with a state-$k%
{\acute{}}%
$ with {\em odd-quasi-parity} when $|\alpha _{o}|=\left( k-k%
{\acute{}}%
\right) /2.$ \ However, when $\alpha _{o}=0$ the central-like core becomes
attractive and the corresponding states cease to perform {\em %
quasi-parity-oscilations. }For more details on the result (27) the reader
may refer to Znojil and L\'{e}vai [21].

\subsection{${\cal PT-}$symmetric harmonic oscillator $V(r)=r^{2}$}

For this potential the leading energy term reads 
\begin{equation}
\bar{l}^{2}E_{k,\ell }^{(-2)}=\frac{\bar{l}^{2}}{r_{0}^{2}}+r_{0}^{2},
\end{equation}
and supports four eligible minima ( all satisfy our conditions in (17))
obtained through $r_{0}^{4}=\bar{l}^{2}$ as $r_{0}=\pm ~i|~\bar{l}~|^{1/2}$
\ and $r_{0}=\pm ~|~\bar{l}~|^{1/2}$. In this case $\Omega =2,$%
\begin{equation}
\beta =\beta ^{(+)}=-~(2k+3/2)
\end{equation}
for \ $r_{0}=\pm ~|~\bar{l}~|^{1/2}$ , and 
\begin{equation}
\beta =\beta ^{(-)}=~(2k+1/2)
\end{equation}
for \ $r_{0}=\pm ~i~|~\bar{l}~|^{1/2}$ . Whilst the former (29) yields 
\begin{equation}
\bar{l}^{2}E_{k,\ell }^{(-2)}=2r_{0}^{2}=4k+2\ell _{d}+3,
\end{equation}
the latter (30) yields 
\begin{equation}
\bar{l}^{2}E_{k,\ell }^{(-2)}=2r_{0}^{2}=4k+1-2\ell _{d}.
\end{equation}
In both cases $\beta =\beta ^{(\pm )}$ the higher-order corrections vanish
identically. Yet, one would combine (31) and (32) by the superscript $(\pm )$
and cast 
\begin{equation}
E_{k,\ell }^{(\pm )}=4k+2\pm (2\ell _{d}+1).
\end{equation}

Therefore, the ${\cal PT-}$ symmetric oscillator is exactly solvable, by our
recipe, and its spectrum, non-equidistant in general, exhibits some unusual
features ( cf. [22] for more details). However, it should be noted that for
the one-dimensional oscillator ( where $\ell _{d}=-1,$ and$~0,$ even and odd
parity, respectively) equation (33) implies (I) $E^{(+)}/2=2k+1/2,$ $%
E^{(-)}/2=2k+3/2$, for $\ell _{d}=-1$, and (II) $E^{(+)}/2=2k+3/2$, $%
E^{(-)}/2=2k+1/2$, for $\ell _{d}=0$.\ Which can be combined together to
form the exact well known result 
\begin{equation}
E_{N}=2N+1;\qquad N=0,1,2,....
\end{equation}
with a new, redefined quantum number $N$.

\section{Application: ${\cal PT-}$symmetric ${\em DDT}$ oscillators}

In our \ ${\cal PT-}$symmetric Schr\"{o}dinger equation (1) with the
practical effective potential 
\begin{equation}
V_{p}(r)=\frac{\ell (\ell +1)}{r^{2}}-\alpha \,\sqrt{~i\,r}+i\,r^{3}
\end{equation}
the general solutions themselves are analytic functions of $r$ (c.f., e.g.,
[6]). We may construct them in the complex plane which is cut, say, from the
origin upwards. This means that $r=\xi \,\exp (i\,\varphi )$ with the length 
$\xi \in (0,\infty )$ and with the span of the angle $\varphi \in (-3\pi
/2,\pi /2)$. Compact accounts of the related mathematics may be found in
Bender et al [8].

Let us proceed with our ${\cal PT}$-PSLET and search for the minimum/minima
of our leading energy term for the {\em DDT}-oscillators (1) 
\begin{equation}
\bar{l}^{2}E_{k,\ell }^{(-2)}=\frac{\bar{l}^{2}}{r_{0}^{2}}-\alpha \sqrt{%
~i\,r_{0}}+i~r_{0}^{3}.
\end{equation}
Evidently, condition (17) yields 
\begin{equation}
r_{0}{}^{5}+i\,\left[ \frac{1}{6}\alpha ~r_{0}{}^{2}\sqrt{~ir_{0}}+\frac{2}{3%
}~\bar{l}^{2}\right] =0.
\end{equation}
Obviously, a closed form solution for this equation is hard to find ( if it
exists at all) and one has to appeal to numerical techniques to solve for $%
r_{0}.$

{\em A priori}, it is convenient to do some elementary analyzes, in the
vicinity of the extremes of $\alpha $ ( mandated by condition (2)), and
distinguish between the two different domains of $\alpha $. For this purpose
let us denote $(2\bar{l}^{2})/3=G^{2}$, re-scale $r_{0}=-~i|G^{2/5}|\,\rho $
and abbreviate $\ \alpha =\delta \sqrt{6~G^{2}}\,\ $ with $0\lesssim $ $%
\delta \lesssim 1$. This gives the following new algebraic transparent form
of our implicit definition of the minimum/minima in (37), 
\begin{equation}
1-Z=\delta ~\sqrt{\frac{Z}{6}}\,\ ;.~~~Z=\rho ^{5}.
\end{equation}

In the weak-coupling domain, vanishing $\delta \approx 0,$ equation (38)
becomes trivial ($1-Z=0)$. It is easy to verify that (36), with $\delta
\approx 0,$ has a {\em unique absolute minimum \ }at{\em \ } 
\begin{equation}
Z=1\Longrightarrow \rho =1\Longrightarrow r_{0}=-~i|G^{2/5}|~,~~\delta =0.
\end{equation}
In the strong-coupling regime, $\delta \approx 1,$ equation (38) yields $%
Z=2/3.$

At this point, one may choose to work with $\beta =0$ ( i.e. $E_{k,\ell
}^{(-1)}\neq 0$) and obtain the leading (zeroth)-order approximation 
\begin{equation}
\bar{l}^{2}E_{k,\ell }^{(-2)}=\left( \frac{\,G^{6}}{Z^{2}}\right) ^{1/5}%
\left[ 5Z-\frac{15}{2}\right] ,\ 
\end{equation}
and, with 
\[
\frac{\omega ^{2}}{4}=B_{2}=\frac{1}{r_{0}^{4}}{\Large (}\frac{5}{2}{\Large )%
}[1+Z],\qquad r_{0}=-iG^{2/5}Z^{1/5},
\]
the first-order correction 
\begin{eqnarray}
\bar{l}E_{k,\ell }^{(-1)} &=&\sqrt{\frac{3}{2}}G{\LARGE \ }\left[ \frac{1}{%
r_{0}^{2}}+(2k+1)\frac{\omega }{2}\right]   \nonumber \\
&=&\left( \frac{G}{Z^{2}}\right) ^{1/5}\left[ \sqrt{\frac{15~(1+Z)}{4}}%
(2k+1)-\sqrt{\frac{3}{2}}\right] .\ 
\end{eqnarray}
Consequently, the energy series (8) reads, up to the first-order correction 
\begin{equation}
E_{k,\ell }=\frac{1}{Z^{2/5}}\left[ \,G^{6/5}~(5Z-\frac{15}{2}%
)+G^{1/5}~\left( \sqrt{\frac{15~(1+Z)}{4}}(2k+1)-\sqrt{\frac{3}{2}}\right) %
\right] .
\end{equation}
Nevertheless, one may choose to work with $\beta =\beta ^{\left( -\right)
}\neq 0$ \ ( i.e. $E_{k,\ell }^{(-1)}=0$) and obtain 
\begin{equation}
\beta =\beta ^{\left( -\right) }=-\frac{1}{2}\left[ 1-(\ 2k+1)\sqrt{5~(1+Z)/2%
}{}\right] .
\end{equation}
Thus the zeroth-order approximation yields 
\begin{equation}
\bar{l}^{2}E_{k,\ell }^{(-2)}=\left( \frac{\,G_{s}^{6}}{Z^{2}}\right) ^{1/5}~%
\left[ 5Z-\frac{15}{2}\right] ~;\ ~G_{s}=\sqrt{\frac{2}{3}}\left[ \ell _{d}+%
\frac{1}{2}-(2k+1)\sqrt{\frac{5~(1+Z)}{8}}\right] .
\end{equation}

In Figure I we plot the energies of (44) vs $\ell \in (-5,~5)$ at different
values of $Z=Z(\delta )\in (2/3,$ $1)$. Obviously, our results show that
even with $\ell <\ell _{DDT}~(\alpha )$ the spectrum remains real and
discrete. Moreover, once we replace $\ell (\ell +1)\longrightarrow $ $\alpha
_{o}^{2}-1/4,$ i.e. $\ell \longrightarrow -1/2+q|\alpha _{o}|$ with $q=\pm 1$
denoting quasi-parity, {\em quasi-parity-oscillations }are manifested by the
unavoidable energy levels crossings, see Figure II.

Table I shows that our results from Eqs.(42) and (44) compare satisfactorily
with those obtained by Znojil, Gemperle and Mustafa [2], via direct variable
representation (DVR). We may mention that even in the domain of not too
large $\ell ,$ the difference between the exact and approximate energies
remain small, of order of $\thickapprox 0.05\%$ from Eq.(42), with $\beta
=0, $ and $\thickapprox 0.2\%$ from Eq.(44), with $\beta =\beta ^{(-)}\neq
0, $ for the ground state. Such a prediction should not mislead us in
connection with the related convergence/divergence of our energy series (8),
which is in fact the genuine test of our present \ ${\cal PT-}$symmetric
PSLET formulae. The energy series (8) with $\beta =\beta ^{(-)}\neq 0$
converge more rapidly than it does with $\beta =0.$ Nevertheless, our
leading energy term remains the benchmark for testing the reality and
discreteness of the energy spectrum.

In Table II we compare our results ( using $\beta =\beta ^{(-)}\neq 0,$
hereinafter, numerically solve for $r_{0}$ and following the procedure of
section III)\ for (1) with $\alpha =0$ using the first ten-terms of (8) and
the corresponding Pad\'{e} approximant, again with those from the DVR
approach. They are in almost exact accord. Hereby, we may emphasize that the
digital-precision enhances for larger $\bar{l}$ ( smaller $1/\bar{l}$ )
values, where the energy series (8) and the related Pad\'{e} approximants
stabilize more rapidly.

Extending the recipe of our test beyond the weak-coupling regime $\delta
\thickapprox 0$ (i.e. $\alpha \thickapprox 0$) we show, in table III, the
energy dependence on the non-vanishing $\alpha .$ Evidently, the
digital-precision of our \ ${\cal PT}$ - PSLET recipe reappears to be $\bar{l%
}$-dependent and almost $\alpha $ - independent. In table IV we witness that
the leading energy approximation inherits a substantial amount of the total
energy documenting, on the computational and practical methodical side, the
usefulness of our pseudo-perturbation recipe beyond its promise of being
quite handy. Yet, a broad range of $\alpha $ is considered including the
domain of negative values, safely protected against any possible spontaneous 
${\cal PT}$ -symmetry breaking.

\section{Summary}

Our purpose was to find a suitable {\em perturbation series like} expansion
of the bound states. We have started from the observation that the physical
consistency of the model (1) (i.e., the reality of its spectrum) is
characterized by the presence of a strongly repulsive/attractive core in the
potential $V_{p}(r)$. This is slightly counterintuitive since the
phenomenologically useful values of $\ell $ are usually small and only the
first few lowest angular momenta are relevant in the Hermitian
Schr\"{o}dinger equations with central symmetry.

Only exceptionally, very high partial waves are really needed for
phenomenological purposes (say, in nuclear physics [23]). The strong
repulsion is required there, first of all, due to its significant {\em %
phenomenological} relevance and {\em in spite} of the formal difficulties.

Fortunately, efficient $\ell \gg 1$ approximation techniques already exist
for the latter particular realistic Hermitian models. They have been
developed by many authors (cf., e.g., their concise review in [24]). Their
thorough and critical tests are amply available but similar studies were
still missing in the non-Hermitian context.

Our present purpose was to fill the gap at least partially. We have paid
thorough attention to the first few open problems related, e.g., to the
possible complex deformation of the axis of coordinates. Our thorough study
of a few particular ${\cal PT}$ symmetric examples revealed that the
transition to the non-Hermitian models is unexpectedly smooth. We did not
encounter any serious difficulties, in spite of many apparent obstacles as
mentioned in Section II (e.g., an enormous ambiguity of the choice of the
most suitable zero-order approximation).

In this way, our present study confirmed that the angular momentum (or
dimension) parameter $\ell $ in the ``strongly spiked'' domain where $|\ell
|\gg 1$ offers its formal re-interpretation and introduction of an
artificial perturbation-like parameter $1/\bar{l}$ which may serve as a
guide to the interpretation of many effective potentials as a suitably
chosen solvable (harmonic oscillator) zero-order approximation followed by
the systematically constructed corrections which prove obtainable quite
easily.

\section*{Acknowledgments}

The participation of MZ in this research was partially supported by GA AS
(Czech Republic), grant Nr. A 104 8004.

\section*{Figure captions}

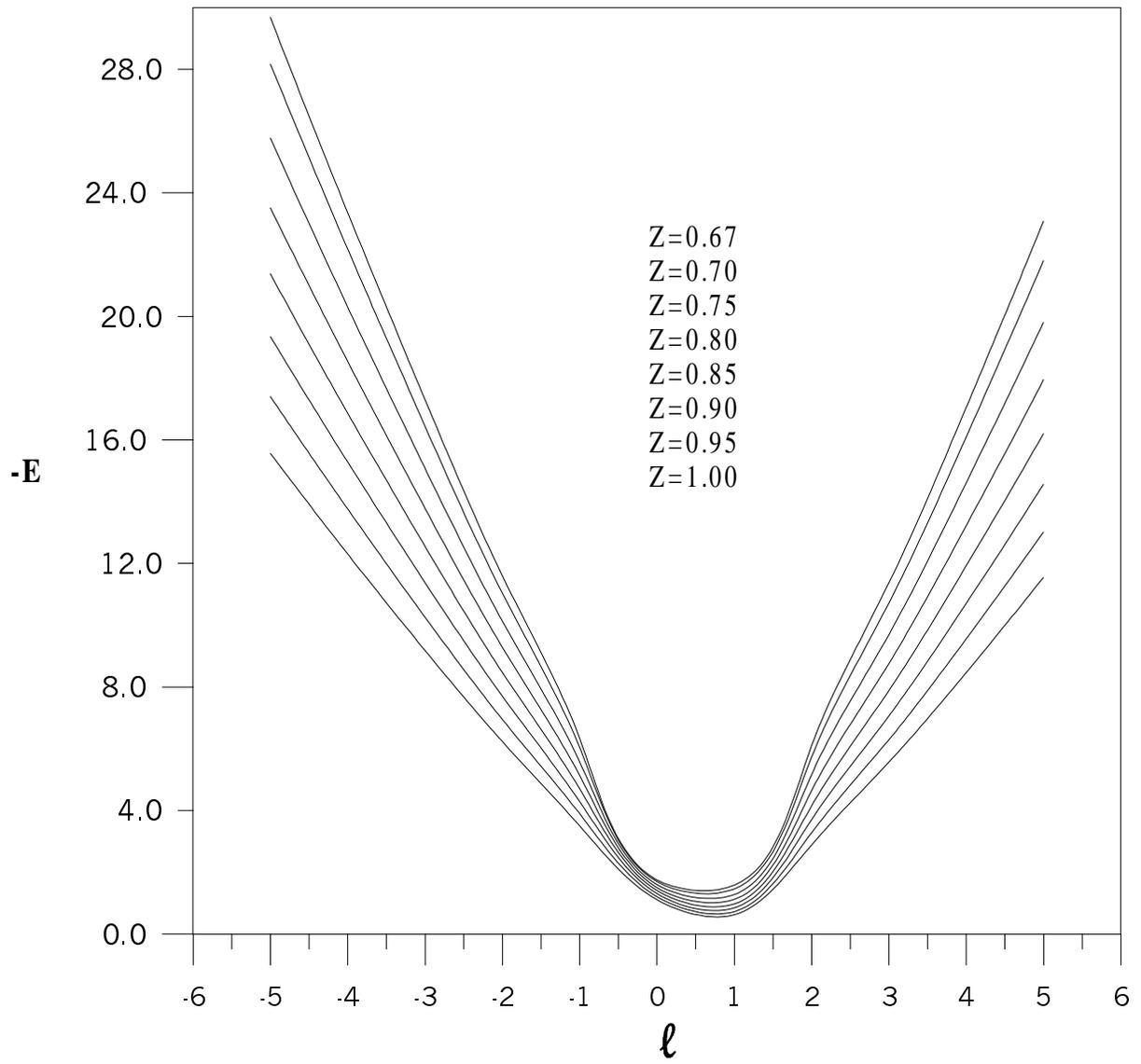
{\bf Figure 1}. Cubic oscillator (35) eigenenergies $(-E)$ in (44) vs $\ell $
for $k=0$ and \ $|~\ell ~|<5$ \ at different values of $Z=Z(\delta )\in
(2/3, $ $1).$

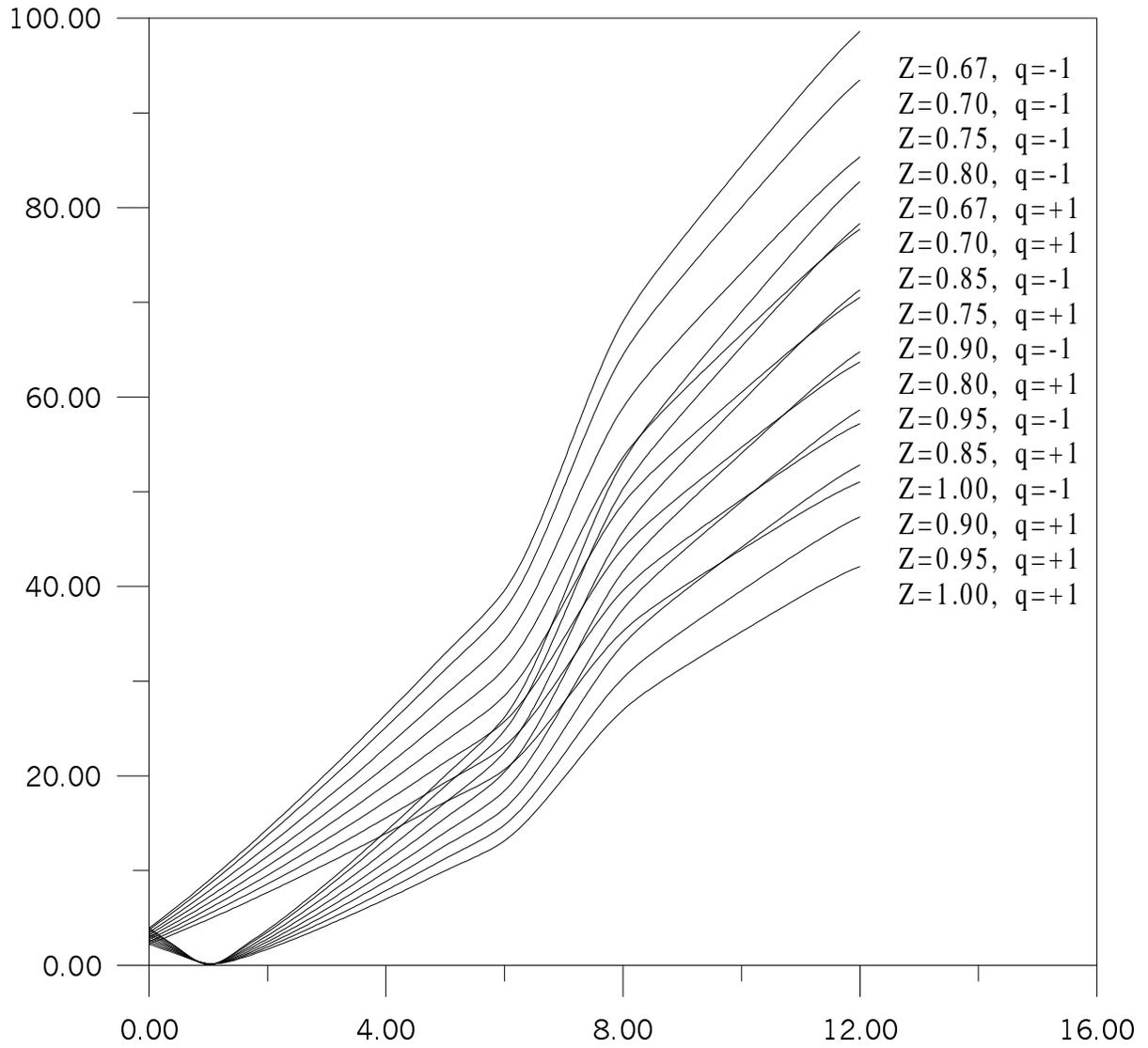
{\bf Figure 2}. Cubic oscillator (35) eigenenergies $(-E)$ in (44) vs $%
|\alpha _{o}|$ for $k=0,$ $\ell =-1/2+q|\alpha _{o}|,$ and different values
of $Z=Z(\delta )\in (2/3,$ $1)$ at {\em even}- and {\em odd-quasi-parities}.

\newpage 
\begin{table}[tbp]
\caption{Comparison of the energy levels for model (1) ( with $\protect\alpha%
=0$). The benchmark, numerically exact DVR values are cited from Znojil,
Gemperle and Mustafa [2].}
\begin{center}
\begin{tabular}{ccccc}
\hline
$\ell$ & $k$ & DVR & Eq.(42) & Eq.(44) \\ \hline
5 & 0 & -11.52191 & -11.517 & -11.542 \\ 
& 1 & -4.56482 & -4.260 & -4.900 \\ 
& 2 & 1.87017 & 2.997 & -0.109 \\ \hline
10 & 0 & -28.76552 & -28.762 & -28.776 \\ 
& 1 & -20.59867 & -20.426 & -20.756 \\ 
& 2 & -12.70640 & -12.090 & -13.230 \\ 
& 3 & -5.11663 & -3.754 & -6.380 \\ 
& 4 & 2.14032 & 4.58 & -0.727 \\ \hline
20 & 0 & -68.72646 & -68.724 & -68.733 \\ 
& 1 & -59.24706 & -59.149 & -59.330 \\ 
& 2 & -49.91773 & -49.574 & -50.171 \\ 
& 3 & -40.74589 & -39.998 & -41.283 \\ 
& 4 & -31.73951 & -30.423 & -32.705 \\ 
& 5 & -22.90712 & -20.847 & -24.489 \\ 
& 6 & -14.25769 & -11.272 & -16.713 \\ 
& 7 & -5.80054 & -1.696 & -9.512 \\ 
& 8 & 2.45491 & 7.879 & -3.172 \\ \hline
50 & 0 & -211.13555 & -211.134 & -211.138 \\ 
& 1 & -199.68009 & -199.633 & -199.718 \\ 
& 2 & -188.29459 & -188.132 & -188.406 \\ 
& 3 & -176.98040 & -176.631 & -177.206 \\ 
& 4 & -165.73889 & -165.129 & -166.123 \\ 
& 5 & -154.57149 & -153.628 & -155.161 \\ 
& 6 & -143.47967 & -142.127 & -144.328 \\ 
& 7 & -132.46494 & -130.626 & -133.628 \\ 
& 8 & -121.52886 & -119.124 & -123.069 \\ \hline
\end{tabular}
\end{center}
\end{table}
\newpage 
\begin{table}[tbp]
\caption{ Same as table I with ${\cal PT-}$PSLET results from the first
ten-terms of (8) and the corresponding Pad\'{e} approximant.}
\begin{center}
\begin{tabular}{cccccc}
\hline
$k$ & $\ell$ & DVR & ${\cal PT-}$PSLET & Pad\'{e} & $\bar{l}\approx$ \\ 
\hline
0 & 5 & -11.52191 & -11.52191 & -11.52191336 & 4.4 \\ 
& 10 & -28.76552 & -28.76552 & -28.76552178 & 9.4 \\ 
& 20 & -68.72646 & -68.72646 & -68.72645928 & 19.4 \\ 
& 50 & -211.13555 & -211.135548 & -211.13554785 & 49.4 \\ \hline
1 & 5 & -4.56482 & -4.565 & -4.564813 & 2.2 \\ 
& 10 & -20.59867 & -20.598669 & -20.59866911 & 7.2 \\ 
& 20 & -59.24706 & -59.24705556 & -59.247055572 & 17.2 \\ 
& 50 & -199.68009 & -199.68008657 & -199.6800865747 & 47.2 \\ \hline
2 & 10 & -12.70640 & -12.7065 & -12.7964017 & 4.9 \\ 
& 20 & -49.91773 & -49.917727 & -49.917727248 & 14.9 \\ 
& 50 & -188.29459 & -188.294591 & -188.29459127507 & 44.9 \\ \hline
3 & 10 & -5.11663 & -9.388 & -5.1166 & 2.7 \\ 
& 20 & -40.74589 & -40.74589 & -40.74589026 & 12.7 \\ 
& 50 & -176.98040 & -176.98039968 & -176.98039968 & 42.7 \\ \hline
\end{tabular}
\end{center}
\end{table}
\newpage 
\begin{table}[tbp]
\caption{Comparison of the energies for model (1) with $\ell=10$ and
different values of $\protect\alpha$.}
\begin{center}
\begin{tabular}{ccccc}
\hline
$\alpha$ & $k$ & DVR & ${\cal PT-}$PSLET & Pad\'{e} \\ \hline
20 & 0 & -58.62190 & -58.6219026 & -58.621902667 \\ 
& 1 & -49.83626 & -49.836258 & -49.8362579 \\ 
& 2 & -41.29014 & -41.2905 & -41.29014 \\ \hline
10 & 0 & -43.85223 & -43.85223324 & -43.852233235 \\ 
& 1 & -35.40717 & -35.407174 & -35.40717408 \\ 
& 2 & -27.23003 & -27.23010 & -27.23003 \\ \hline
0 & 0 & -28.76552 & -28.76552 & -28.76551777 \\ 
& 1 & -20.59867 & -20.59867 & -20.598669108 \\ 
& 2 & -12.70640 & -12.70653 & -12.7064017 \\ \hline
-10 & 0 & -13.35529 & -13.3552878 & -13.355287796 \\ 
& 1 & -5.40717 & -5.40717 & -5.4071736 \\ 
& 2 & 2.27617 & 2.08 & 2.27617 \\ \hline
-20 & 0 & 2.38131 & 2.381306 & 2.38130557 \\ 
& 1 & 10.16462 & 10.1646 & 10.1646191 \\ 
& 2 & 17.70339 & ... & ... \\ \hline
\end{tabular}
\end{center}
\end{table}
\newpage 
\begin{table}[tbp]
\caption{Energies for model (1) with $\ell=0$ and different values of $%
\protect\alpha, d, k$.}
\begin{center}
\begin{tabular}{ccccccc}
\hline
$d$ & $\alpha$ & $k$ & Leading term & ${\cal PT-}$PSLET & Pad\'{e} & $\bar{l}%
\approx$ \\ \hline
3 & -5 & 0 & 3.07 & 2.903 & 2.881 & 1 \\ 
&  & 1 & -1.65 & -1.58 & -1.579391 & 3.2 \\ 
&  & 2 & -7.962 & -7.652 & -7.65505 & 5.4 \\ \hline
& -10 & 0 & 8.23 & 7.82 & 7.820 & 1.3 \\ 
&  & 1 & 3.98 & 3.803 & 3.80992 & 3.5 \\ 
&  & 2 & -1.87 & -1.80 & -1.798903 & 5.7 \\ \hline
& -15 & 0 & 13.95 & 13.38 & 13.38419 & 1.6 \\ 
&  & 1 & 9.95 & 9.52 & 9.542833 & 3.8 \\ 
&  & 2 & 4.45 & 4.27 & 4.273163 & 6.0 \\ \hline
1 & 10 & 0 & -12.46 & -12.471 & -12.471 & 1.4 \\ 
&  & 1 & -20.65 & -20.304 & -20.3040 & 3.5 \\ 
&  & 2 & -29.00 & 028.341 & -28.34038 & 5.7 \\ \hline
& 5 & 0 & -8.12 & -8.082 & -8.082 & 1.5 \\ 
&  & 1 & -15.38 & -15.058 & -15.05795 & 3.6 \\ 
&  & 2 & -23.15 & -22.569 & -22.56820 & 5.9 \\ \hline
& -5 & 0 & 1.57 & 1.5393 & 1.5393 & 1.8 \\ 
&  & 1 & -4.16 & -4.055 & -4.055465 & 4.0 \\ 
&  & 2 & -10.94 & -10.63 & -10.6341055 & 6.3 \\ \hline
& -15 & 0 & 12.99 & 12.754 & 12.75391 & 2.2 \\ 
&  & 1 & 8.06 & 7.84 & 7.84006 & 4.6 \\ 
&  & 2 & 1.99 & 1.93 & 1.9269341 & 6.8 \\ \hline
\end{tabular}
\end{center}
\end{table}

\end{document}